\documentclass[twocolumn,showpacs,preprintnumbers,amsmath]{revtex4}
\usepackage{amssymb}

\usepackage{graphicx}
\usepackage{dcolumn}
\usepackage{bm}

\input{tcilatex}
\begin{document}

\title{Time-dependent local Green's operator \\
and its applications to manganites}
\author{H. Aliaga}
\date{\today}

\begin{abstract}
An algorithm is presented to calculate the electronic local time-dependent
Green's operator for manganites-related hamiltonians. This algorithm is
proved to scale with the number of states $N$ in the Hilbert-space to the
1.55 power, is able of parallel implementation, and outperforms
computationally the Exact Diagonalization (ED) method for clusters larger
than 64 sites (using parallelization). This method together with the Monte
Carlo (MC) technique is used to derive new results for the manganites phase
diagram for the spatial dimension D=3 and half-filling on a 12x12x12 cluster
(3456 orbitals). We obtain as a function of an insulating parameter, the
sequence of ground states given by: ferromagnetic (FM), antiferromagnetic
AF-type A, AF-type CE, dimer and AF-type G, which are in remarkable
agreement with experimental results.
\end{abstract}

\maketitle

\preprint{APS/123-QED}


\affiliation{Lehrstuhl f\"{u}r Theoretische Physik III, Institut f\"{u}r
Physik, Universit\"{a}t Augsburg,D-86135 Augsburg, Germany. }



\section{Introduction}

The doped manganites are interesting not only because of the potential
applications of their 'colossal' properties, but also because they are
fascinating systems to study due to the delicate balance of interactions
between charge, spin, and orbital degrees of freedom.\cite{kaplan}

One of the successful theoretical approaches to the study of the manganites
phase diagrams, is based on effective models that consider the competition
between double-exchange (DE), superexchange (SE), and the electron-phonon
interactions\cite{1}. These models and techniques have predicted interesting
phenomena, like nanoscale phase-separation\cite{2}, influence of disorder in
metal-insulator transitions\cite{3}, and the existence of non-trivial
magnetic phases\cite{4}, among other phenomena. The ground states of such
systems and the finite-temperature properties were obtained using
simulations on clusters of spins using the Monte Carlo (MC) technique
together with the calculation of the electronic energies by means of the
exact diagonalization (ED) technique\cite{2} at each single MC step. In
order to extract useful information from the simulations, one needs to
analyze different sizes for the clusters, where the CPU (Central Processing
Unit) memory and time scale as the number of states $N$ in the Hilbert space
to the third- and fourth-power, respectively. In 1999, the Truncated
Polynomial Expansion (TPEM)\cite{motome1} was devised in order to reduce the
CPU time scaling with $N$, but unfortunately at a cost of a large prefactor
and a detailed comparison with exact case (ED) was not shown.

Moreover, it is of fundamental importance to consider the Mn-$e_{g}$
orbitals $d_{x^{2}-y^{2}}$ and $d_{3z^{2}-r^{2}}$ at each site\cite{kaplan},
which doubles the size of the matrices respect of the case of a single band.
The ED operations have to be performed several thousands times in a typical
simulation, practically limiting the spatial dimensions of the clusters
considered to D=1 and 2. The reduction in the dimensionality of the system
artificially breaks the degeneracy of the orbitals $d_{x^{2}-y^{2}}$ and $%
d_{3z^{2}-r^{2}}$, modifying the electronic properties of the system.

For all these reasons it is important to find an alternative way to
calculate the electronic energy that avoids the natural limitations of the
ED scheme, allowing to expand the size of the clusters and the spatial
dimensionality. An efficient and fast way to calculate the time-dependent
Green's operator in an effective model conceived for manganites will be
described in this work. This method uses the Chebyshev expansion of the time
evolution operator\cite{5} which is an extremely accurate and fast way to
obtain the dynamics response of quantum many-body systems. For a system
consisting of 10 spins S=1/2 with Heisenberg-like interactions, this
expansion is three orders of magnitude faster than the ED method, and for a
variety of considered cases, is one of the most efficient time-marching
schemes to keep track of the time evolution of a quantum system, where CPU
memory and time basically scale linearly with system size.\cite{5}

In this work we will focus our attention to non-trivial single-particle
Hamiltonians, where the Hilbert space grows linearly with the size of the
system. More generally speaking, the presently described method, which will
be referred as the \textit{Dyn }method\textit{,} provides an alternative way
to calculate the local time-dependent Green's operator for the
time-dependent Schr\"{o}dinger equation (TDSE). The Chebyshev expansion of
the hermitian, time-independent, differential Hamiltonian operator $H$ will
be applied to a model related to manganites, which is more efficient than
the ED technique for electronic clusters larger than 64 orbitals (using full
parallelization). The feasibility of the method will be tested proposing
fixed spins configuration and comparing the densities of states with well
known analytic results. Then the ground states, electronic properties and
phase diagram for D=3 will be determined by means of the \textit{Dyn} method
together with the MC technique and a comparison with experimental results
will be made.

The text is organized as follows: in Section II the Hamiltonian model is
described, the formalism regarding the calculation of the time-dependent
local Green's operator is given in Section III and in Section IV a benchmark
test is used to compare the CPU time as a function of the lattice sites
needed by the ED, TPEM and \textit{Dyn} methods. In Sections V and VI, the
cases of one orbital per site are considered. For fixed configurations of
the spins, the electronic density of states are calculated and compared with
analytical results, and the relative error is discussed. In Section VII the 
\textit{Dyn} method is used together with the Monte Carlo technique to
obtain the ground state and reproduce the phase diagram in two spatial
dimensions (D=2), and the results are compared with the MC+ED technique. In
Section VIII the MC+\textit{Dyn} technique is used to obtain the phase
diagram for filling x=1/2 and D=3 on a 12x12x12 cluster, and the new results
are discussed. Finally, Section IX contains a discussion of the the main
results and general conclusions.

\section{Manganite Effective Hamiltonian}

We will consider throughout this work the cases of (A) one and (B) two
orbitals per site, together with the spatial dimensions, D=1, 2 and 3. The
Hamiltonian model to be considered\cite{4} is quadratic in fermionic
operators, 
\begin{eqnarray}
H &=&-\sum_{\mathbf{{ia}\gamma \gamma ^{\prime }\sigma }}t_{\gamma \gamma
^{\prime }}^{\mathbf{\xi }}(d_{\mathbf{{i}\gamma \sigma }}^{\dag }d_{\mathbf{%
{i+a}\gamma ^{\prime }\sigma }}+h.c.)-J_{\mathrm{H}}\sum_{\mathbf{i}}\mathbf{%
{s}_{{i}}\cdot S}_{{i}}  \nonumber \\
&+&J_{\mathrm{AF}}\sum_{\langle \mathbf{{i,j}\rangle }}\mathbf{{S}_{{i}%
}\cdot {S}_{{j}}+\lambda \sum_{{i}}(}Q_{\mathrm{1}i}\rho _{{i}}+Q_{\mathrm{{2%
}{i}}}\tau _{\mathrm{{x}{i}}}+Q_{\mathrm{{3}{i}}}\tau _{\mathrm{{z}{i}}}) 
\nonumber \\
&+&(1/2)\sum_{\mathbf{i}}(\beta _{r}Q_{1\mathbf{i}}^{2}+Q_{2\mathbf{i}%
}^{2}+Q_{3\mathbf{i}}^{2}),
\end{eqnarray}

For the case (B), the operators $d_{\mathbf{{i}\mathrm{{a}\sigma }}}$ ($d_{%
\mathbf{{i}\mathrm{{b}\sigma }}}$) represent the annihilation of an $e_{%
\mathrm{g}}$-electron with spin $\sigma $, in the $d_{x^{2}-y^{2}}$ ($%
d_{3z^{2}-r^{2}}$) orbital at site $\mathbf{i}$, and $\mathbf{\xi }$ is the
vector connecting nearest-neighboring (NN) sites. The first term in $H$ is
the NN hopping of $e_{\mathrm{g}}$ electrons with amplitude $t_{\gamma
\gamma ^{\prime }}^{\mathbf{\xi }}$ between $\gamma $- and $\gamma ^{\prime
} $-orbitals. For the (B) D=2 case, the hopping amplitude along the $\mathbf{%
\xi }$-direction is given by: $t_{\mathrm{aa}}^{\mathbf{x}}$=$-\sqrt{3}t_{%
\mathrm{ab}}^{\mathbf{x}}$=$-\sqrt{3}t_{\mathrm{ba}}^{\mathbf{x}}$=$3t_{%
\mathrm{bb}}^{\mathbf{x}}$=$t_{ij}t_{h}$, and $t_{\mathrm{aa}}^{\mathbf{y}}$=%
$\sqrt{3}t_{\mathrm{ab}}^{\mathbf{y}}$=$\sqrt{3}t_{\mathrm{ba}}^{\mathbf{y}}$%
=$3t_{\mathrm{bb}}^{\mathbf{y}}$=$t_{ij}t_{h}$. The parameter $t_{h}$ is the
hopping transfer integral and will from now on the energy unit of the model,
and the parameter $t_{ij}$ is a complex scalar that depends on the relative
orientation of neighboring localized spins $\mathbf{S}_{i}$ and $\mathbf{S}%
_{j}$, assumed classical with $|\mathbf{{S}|}$=1, and characterized by the
polar and azimutal angles, $\theta _{i}$ and $\varphi _{i}$\cite{2}

\begin{equation}
t_{ij}=\cos (\theta _{i}/2)\cos (\theta _{j}/2)+e^{-i(\varphi _{i}-\varphi
_{j})}\sin (\theta _{i}/2)\sin (\theta _{j}/2)
\end{equation}
when the approximation of very large $J_{\mathrm{H}}$ is used. It should be
kept in mind that for $\lambda =0$ the original degeneracy of the $%
d_{x^{2}-y^{2}}$ and $d_{3z^{2}-r^{2}}$ orbitals is explicitly broken in the
D=2 case. The hopping process along the $\mathbf{x}$- and $\mathbf{y}$-
directions favor the occupation of the $d_{x^{2}-y^{2}}$ over the $%
d_{3z^{2}-r^{2}}$ orbitals in an important range of fillings. For that
reason it is important to consider the D=3 case, where the degeneracy is
recovered when the hopping processes along the $\mathbf{a}$=$\mathbf{z}$
direction are included, that is for $t_{\mathrm{aa}}^{\mathbf{z}}$=$t_{%
\mathrm{ab}}^{\mathbf{z}}$=$t_{\mathrm{ba}}^{\mathbf{z}}$=0 and $t_{\mathrm{%
bb}}^{\mathbf{z}}$=$4t_{ij}t_{h}/3$.

In the second term of Eq. (1), the Hund constant $J_{\mathrm{H}}$($>$0)
couples the spin $\mathbf{{s}_{{i}}}$= $\sum_{\gamma \nu _{1}\nu _{2}}d_{%
\mathbf{{i}\gamma \nu _{1}}}^{\dag }\mathbf{\sigma }_{\nu _{1}\nu
_{2}}d_{i\gamma \nu _{2}}$ ($\mathbf{\sigma }$=Pauli matrices) of the $e_{%
\mathrm{g}}$ electrons with the localized $t_{\mathrm{2g}}$-spin $\mathbf{{S}%
_{{i}}}.$ The constant $J_{\mathrm{H}}$ is here considered as infinite or
very large, and the model is drastically simplified.\cite{2}

The third term is the AF coupling $J_{\mathrm{AF}}$ between NN $t_{\mathrm{2g%
}}$ spins. The fourth term couples $e_{\mathrm{g}}$ electrons and MnO$_{6}$
octahedral distortions, $\lambda $ is a dimensionless coupling constant, $%
Q_{1\mathbf{i}}$ is the breathing-mode distortion, and $Q_{2\mathbf{i}}$ and 
$Q_{3\mathbf{i}}$ are, respectively, the $(x^{2}$$-$$y^{2})$- and $(3z^{2}$$%
- $$r^{2})$-type Jahn-Teller ($JT$) mode distortions, $\rho _{\mathbf{i}}$= $%
\sum_{\gamma ,\sigma }d_{\mathbf{{i}\gamma \sigma }}^{\dag }d_{\mathbf{{i}%
\gamma \sigma }}$, $\tau _{\mathrm{{x}\mathbf{i}}}$= $\sum_{\sigma }(d_{%
\mathbf{{i}\mathrm{{a}\sigma }}}^{\dag }d_{\mathbf{{i}\mathrm{{b}\sigma }}}$
+$d_{\mathbf{{i}\mathrm{{b}\sigma }}}^{\dag }d_{\mathbf{{i}\mathrm{{a}\sigma 
}}})$, and $\tau _{\mathrm{{z}\mathbf{i}}}$= $\sum_{\sigma }(d_{\mathbf{{i}%
a\sigma }}^{\dag }d_{\mathbf{{i}a\sigma }}$ $-$$d_{\mathbf{{i}b\sigma }%
}^{\dag }d_{\mathbf{{i}b\sigma }})$. The fifth term is the quadratic
potential for adiabatic distortions and $\beta _{r}$ is the ratio of spring
constants for breathing- and $JT$-modes, which for manganites is
approximately given by $\beta _{r}\approx $2.\cite{hotta}

In the (A) case, one spherical $s$ orbital per site will be considered and
the orbital indexes in Eq. (1) can be dropped, that is $\gamma =\gamma
^{\prime }$. The hoppings amplitudes for the different spatial cases are
defined for (A) D=1 as $t^{\mathbf{x}}$=$t_{ij}t_{h}$; for (A) D=2 as $t^{%
\mathbf{x}}$=$t^{\mathbf{y}}$=$t_{ij}t_{h}$ and for (A) D=3 as $t^{\mathbf{x}%
}$=$t^{\mathbf{y}}$=$t^{\mathbf{z}}$=$t_{ij}t_{h}$. For the case (A) the
electron-phonon interaction and the oxygen degrees of freedom are not
considered, that is $Q_{\mathrm{{1}{i}}}$\textit{=}$Q_{\mathrm{{2}{i}}}$%
\textbf{\textit{=}}$Q_{\mathrm{{3}{i}}}$\textbf{\textit{=}}$\lambda $=$\beta
_{r}$=0. This case will allow to compare the density of states obtained with
the $Dyn$ method and well-known analytical results.\cite{Economou}

\section{Time-dependent local Green's operator}

Without loss of generality, the bra and ket notation will be used from now
on. Although the formalism given in this work is oriented to analyze a model
related to manganites (Eq.1), the results are valid for any hermitian,
time-independent, differential operator $H$ expressed in matrix form, which
possesses a complete set of eigenfunctions $\left\{ \mid \phi _{n}\rangle
\right\} $ and eigenvalues $\left\{ \gamma _{n}\right\} $, satisfying $N$
equations of the form,

\begin{equation}
H\mid \phi _{n}\rangle =\gamma _{n}\mid \phi _{n}\rangle  \label{3}
\end{equation}

In case the set of eigenfunctions and eigenvalues are known, the Green's
operator for a one-particle Hamiltonian\cite{Economou} $G(t)$ can be
calculated as,

\begin{equation}
G(t)=-ie^{-iHt/\hbar }=-i\sum_{n=1}^{N}e^{-i\gamma _{n}t/\hbar }\mid \phi
_{n}\rangle \langle \phi _{n}\mid   \label{4}
\end{equation}
where the Planck constant will be set to $\hbar =1$. We can define $%
G^{+}(t)=\Theta \left( t\right) G(t)$, where $\Theta \left( t\right) $ is
the Heaviside step function. After Fourier-transforming $G^{+}(t)$, the
frecuency-dependent Green's operator is obtained,

\begin{equation}
G^{+}(\omega )=\int_{-\infty }^{+\infty }\frac{e^{i\omega t}G^{+}(t)dt}{%
\sqrt{2\pi }}  \label{6}
\end{equation}
which in turn, allows to obtain the electronic density of states,

\begin{equation}
\rho (\omega )=-\frac{1}{\pi } Im \left[ Tr\left( G^{+}(\omega
)\right)\right]  \label{7}
\end{equation}

However, in order to evaluate (Eq.1-6) the knowledge of the eigenfunctions
is needed, which is a difficult task even for non-trivial single particle
Hamiltonians. In this work a novel way is proposed to calculate this local
time-dependent Green's operator, without the explicit knowledge of the set
of eigenfunctions $\left\{ \mid \phi _{n}\rangle \right\} $. It will be
shown later (Fig. 1) that this algorithm outperforms the ED technique for
lattices larger than 64 sites (using parallelization), \textit{without}
providing the information regarding the eigenfunctions $\left\{ \mid \phi
_{n}\rangle \right\} $. The local time-dependent Green's operator must be
expressed in terms of the local basis states $\left\{ \mid n\rangle \right\} 
$ instead of the eigenfunction's set $\left\{ \mid \phi _{n}\rangle \right\} 
$ as

\begin{equation}
G(t)=-i\sum_{n=1}^{N}e^{-iHt}\mid n\rangle \langle n\mid  \label{8}
\end{equation}
and then follow the steps of Eqs. 5-7 to obtain $\rho (\omega )$. It has
been shown by Dobrovitski et al.\cite{5} that using a Chebyshev expansion
method can be an extremely precise and fast way to evaluate the time
evolution operator. The time-dependent Green's functions are matrix elements
of the operator $G,$ and are evaluated as

\begin{equation}
G(i_{0},j_{0};t)=-i\sum_{n=1}^{N}\langle j_{0}\mid n\rangle \langle n\mid
e^{-iHt}\mid i_{0}\rangle  \label{9}
\end{equation}
where $\mid i_{0}\rangle $ and $\mid j_{0}\rangle \in \left\{ \mid n\rangle
\right\} ,$ and the diagonal elements $G(i_{0},i_{0};t)$ are relevant for
determining $\rho (\omega ).$ The quantity $\mid G(i_{0},j_{0};t)\mid ^{2}$
is the probability of creating the particle at site $i_{0}$ and detecting it
at site $j_{0}$, at time a $t$ later, and a decaying behavior with time is
expected.

In order to carry out this expansion, it is first necessary to normalize $H$%
, by a value $\mid \gamma _{\max }\mid $ equal or higher than the highest
eigenvalue in absolute value: $X=H/\mid \gamma _{\max }\mid $. As the $Ht$
phase must be kept constant, the time $t$ is normalized too, by $\tau =t\mid
\gamma _{\max }\mid $.\cite{pick}

Now the expansion of the normalized Hamiltonian $X$ operator at time $\tau $
is,

\begin{equation}
\langle j_{0}\mid e^{-iHt}\mid i_{0}\rangle =\langle j_{0}\mid \left[
J_{0}\left( \tau \right) \mid \nu _{0}\rangle +2\sum_{k=1}^{\infty
}J_{k}\left( \tau \right) \mid \nu _{k}\rangle \right]  \label{10}
\end{equation}

where

\begin{equation}
J_{k}\left( \tau \right) =\int_{-1}^{+1}\frac{e^{-ixt}T_{k}(x)dx}{\sqrt{%
1-x^{2}}}  \label{11}
\end{equation}
are the $k$-order Bessel function of the first kind and $T_{k}(x)$ are the $%
k $-order Chebyshev polynomials of the first kind, given by: $%
T_{k}(x)=\arccos (k\cos (x))$. The vectors $\mid \nu _{k}\rangle $ are
calculated following the Chebyshev recursion expression: $\mid \nu
_{0}\rangle =1\cdot \mid i_{0}\rangle $, $\mid \nu _{1}\rangle =X\cdot \mid
i_{0}\rangle $, and $\mid \nu _{k}\rangle =X\mid \nu _{k-1}\rangle -\mid \nu
_{k-2}\rangle $ (for $k\geqslant 2$). Since the value of a Bessel function
decreases as $J_{k}\left( \tau \right) \approx \left( \tau /k\right) ^{k}$,
the truncation of the series to the order $K_{max}$ leads to an error that
decreases exponentially with $K_{max}$. In practice, holding terms of the
order $K_{max}\approx 1.5\tau $ is enough to get an accuracy of $10^{-7}$ in
the wave function.\cite{accuracy}

\section{Benchmark testing}

The performance of the \textit{Dyn} algorithm was benchmark tested on AMD
Athlon 2500+ (1.85Ghz) processors against the ED algorithm and TPEM\cite
{motome} on square random hermitian matrices. In Fig. 1 are shown the
wall-clock computational times $t$ required to compute Eq. (5) in one Monte
Carlo step per site as a function of the rank of the matrix $H$, $R_{H}$,
using the $ZHEEV-LAPACK$ library (ED algorithm, triangles), TPEM (squares)
and the \textit{Dyn} algorithm (circles). The computational times $t$ were
fitted in the range $100\leq R_{H}\leq 1000$, obtaining the following
dependencies: $log(t_{ED})=-5.47+3.51\log (R_{H}),$ $\log
(t_{TPEM})=-4.59+2.05\log (R_{H})$ and $\log (t_{Dyn})=-3.15+1.55\log (R_{H})
$. For the TPEM and \textit{Dyn} cases the full parallelization capability
was used and $K_{max}=60$ was considered.

\begin{figure}[h]
\includegraphics[width=8cm]{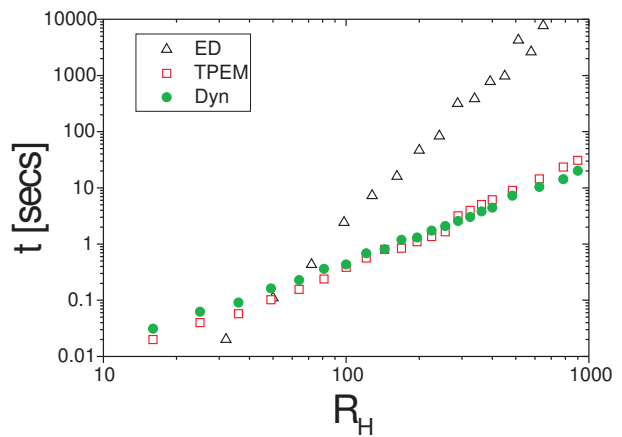}
\caption{Computational time $t$ required to evaluate Eq. (5) in one Monte
Carlo step per site as a function of the rank of the matrix $H$, $R_{H}$,
using ED ($ZHEEV-LAPACK$ library, triangles), TPEM (squares) and \textit{Dyn}
(circles). Full parallelization capability was used for the TPEM and \textit{%
Dyn} codes.}
\label{fig:figure1}
\end{figure}

We can observe that for $R_{H}\gtrsim 64$, the present method and TPEM
outperform the ED technique. For matrices smaller than 144x144, the TPEM
technique is faster than \textit{Dyn, }but this situation is reverted for
matrices of size $R_{H}\geq 289$. This extraordinary performance can be
understood intuitively by the following reasons: 1) The efficience of the
recursive relation of the $\mid \nu _{k}\rangle $ vectors, 2) The functions $%
J_{k}\left( \tau \right) $ are relatively inexpensive to obtain
computationally\cite{motome}, 3) No explicit multiplication between the $%
T_{k}(x)$ operators and the vectors $\mid i_{0}\rangle $ is necessary.\cite
{noneed} 4) The use of the parallelization capability. If only one CPU is
used (no parallelization) in the Dyn (TPEM) algorithm, an increase in a
factor of $R_{H}$ is obtained in the time consumption $t_{Dyn}$ ($t_{TPEM}$%
). In this case \textit{Dyn} outperforms ED only for $R_{H}\gtrsim 900$.

\section{Case A: one orbital per site and FM configuration}

The \textit{Dyn} method can be directly tested with the analytical results
for the density of states in cubic lattices for D=1, 2 and 3. This can be
done by considering one orbital per site for Eq. (1), and the case where $%
\theta _{i}$ and $\varphi _{i}$ are constant for all $i$, that is the
ferromagnetic (FM) phase where all the spins $\mathbf{{S}_{{i}}}$ are
parallel. The effect of the phonons will not be considered and $H$ is
normalized by the value $\mid \gamma _{\max }\mid =2Dt_{h}$.

The \textit{Dyn} method basically is an efficient way to keep track of the
time evolution of a wave-packet. In Fig. 2 can be seen the probability $\mid
G(i_{0},j_{0};\tau )\mid ^{2}$ as a function of the site $j_{0}$, when a
particle is created at time $\tau =0$, in the 500$th$ site of a 1000-site
chain and destroyed at a time $\tau $ later, considering periodic boundary
conditions (PBC). In Fig. 2(a) the snapshot was taken at a time $\tau =1000$%
, where the wave-front is moving away from the $i_{0}$ site, indicated by
the arrows. In the case 2(b) the wave-front has crossed the boundaries of
the chain, and is moving towards the center of the chain, at a time $\tau
=3600$.\cite{feiguin}

It is possible to recover results for the infinite-size limit, provided the
maximum time while the simulation $\tau _{sim}$ is carried out is less than
the time that the 'wave front' $\tau _{wf}$ reaches back to the site $i_{0}$%
, $\tau _{sim}<\tau _{wf}$. In the opposite limit, $\tau _{sim}\gg \tau
_{wf} $, this method obtains the discrete resolution of the density of
states spectra corresponding to a finite-lattice model.

\begin{figure}[h]
\includegraphics[width=8cm]{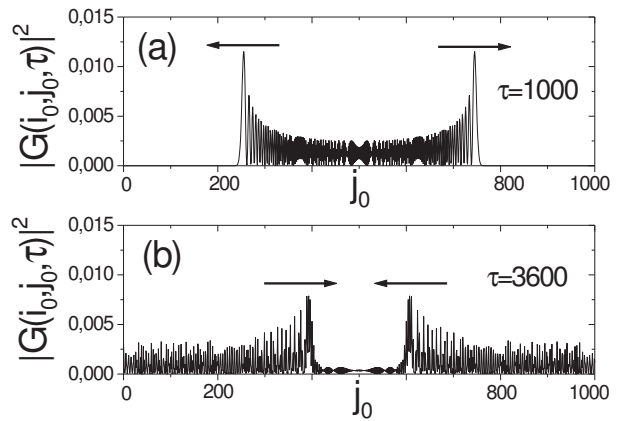}
\caption{ The probability $\mid G(i_{0},j_{0};\tau )\mid ^{2}$ as a function
of the site $j_{0}$, when a particle is created at time $\tau =0$, in the
middle of a 1000-site chain (PBC), and measured at a time $\tau =1000$ later
(a), and at a time $\tau =3000$ (b). The arrows show the direction of
movement of the wave-front. }
\label{fig:figure2}
\end{figure}

\begin{figure}[h]
\includegraphics[width=8cm]{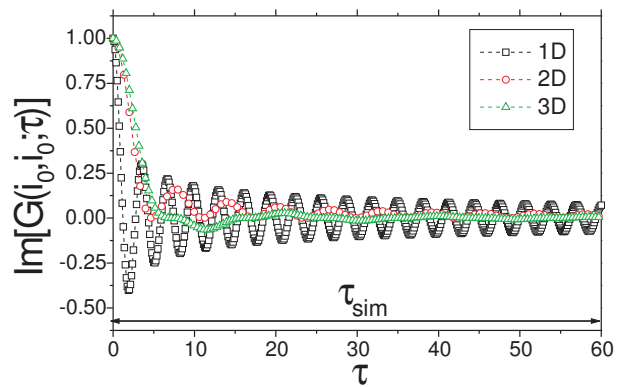}
\caption{Time-dependence of $Im\left[ G(i_{0},i_{0};\tau )\right] $ for the
D=1 (black squares), D=2 (red circles) and D=3 (green triangles) cases, for
lattice sites 1000, 100x100, and 50x50x50, respectively. The simulation is
carried out up to the cut-off time $\tau _{sim}$. At time $\tau =0$, $%
Im\left[ G(i_{0},i_{0};0)\right] $ =1, for D=1, 2 and 3.}
\label{1}
\end{figure}

Keeping track of the wave amplitude at the site $i_{0},$ will allow us to
study the behavior of $G(i_{0},i_{0};\tau )$ for the different spatial
dimensions. In Fig. 3 it is shown the time-dependence of $Im\left[
G(i_{0},i_{0};\tau )\right] $ for the D=1 (black squares, 1000 sites), D=2
(red circles, 100x100 sites) and D=3 (green triangles, 50x50x50 sites)
cases. The corresponding real component of $G(i_{0},i_{0};\tau )$ is zero.

\begin{figure}[h]
\includegraphics[width=8cm]{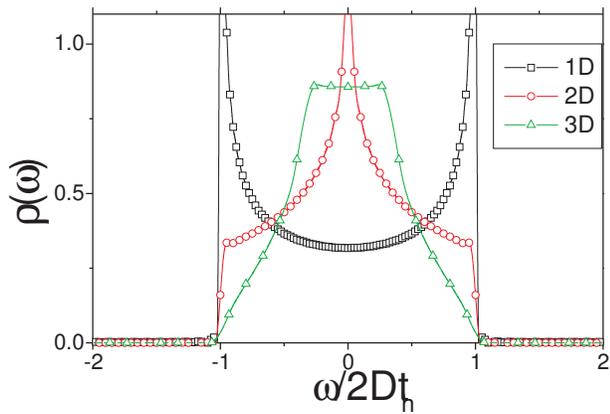}
\caption{ Frequency-dependence of $\rho \left( \omega \right) $ for the FM
case in 1D (black squares), 2D (red circles) and 3D (green triangles).}
\label{fig:figure4}
\end{figure}

After Fourier-transforming the value of $G(i_{0},i_{0};\tau )$, the
corresponding densities of states $\rho \left( \omega \right) $ are obtained
and shown in Fig. 4, and are in remarkably good agreement with analytical
densities of states.\cite{Economou} The simulations are carried out for the
case $\tau _{sim}<\tau _{wf}$, where an approximation to $\rho \left( \omega
\right) $ for an infinite system is obtained.

\begin{figure}[h]
\includegraphics[width=8cm]{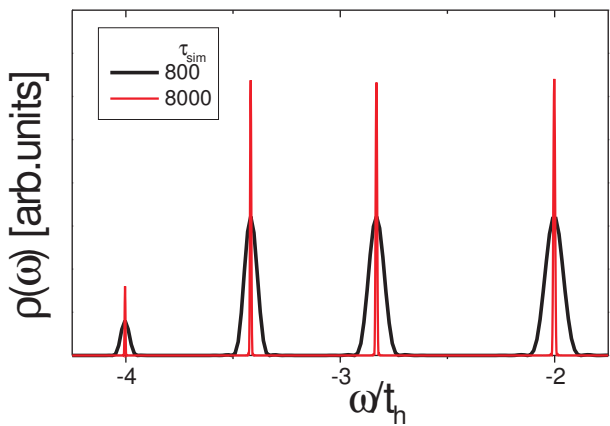}
\caption{Electronic density $\rho \left( \omega \right) $ of states for a
8x8 cluster for the FM case in the energy interval [-4.25$t_{h}$,-1.75$t_{h}$%
], obtained using the\textit{\ Dyn} method, with $\tau _{sim}$=800 (black
line) and $\tau _{sim}$=8000 (red line). Details of the complete spectra
found following the ED method are described in the text. }
\label{fig:figure5}
\end{figure}

In order to compare the discrete nature of the density of states of finite
clusters, a comparison of $\rho \left( \omega \right) $ for an 8x8 cluster
of sites is given in Fig. 5, with $\tau _{sim}$=800 (black line), and $\tau
_{sim}$=8000 (red line). The values picked for $\tau _{sim}$ correspond to
the case $\tau _{sim}\gg \tau _{wf}$, and the greater the value of $\tau
_{sim}$ the higher and narrower are the peaks obtained for $\rho \left(
\omega \right) $, for that reason the distributions are not plotted on the
same vertical scale. The distribution of eigenvalues obtained with ED is as
follows (in units of $t_{h}$): $\pm $4, $\pm $3.41421, $\pm $2.82843, $\pm $%
2, $\pm $1.4142, $\pm $0.58579, and 0. The eigenvalues $\pm $4 are not
degenerate, and the rest of them are four-fold degenerate with the exception
of the eigenvalues $\pm $1.4142 (8 times), and the eigenvalue 0 (14 times).
The peak positions obtained with \textit{Dyn} are in good agreement with the
ED results, where the precision is proportional to $\tau _{sim}$.

The finiteness of the time simulation $\tau _{sim}$ causes an oscillatory
dependence in $\rho \left( \omega \right) ,$ and for values of $\omega $
close to the eigenvalues, $\rho \left( \omega \right) $ can have even small
negative values. This behavior is commonly known as 'overshooting' or Gibbs
oscillations in the Fourier transformation context and these effects are
treated here following the Kernel Polynomial approximation\cite{silver}. The
peaks distributions obtained with $Dyn$ have a finite-frecuency width $%
\Delta \omega \approx \left( 4D\pi \right) /\tau _{sim}$, where $4D$ is the
electronic band-width in units of $t_{h}$.

\begin{figure}[h]
\includegraphics[width=8cm]{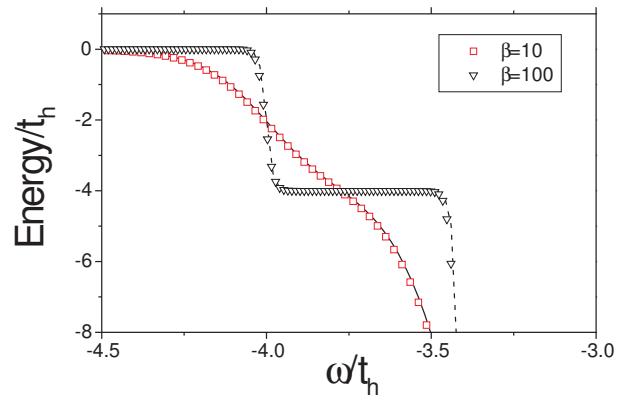}
\caption{Electronic Energy vs. chemical potential. The full (red squares)
and dashed lines (black triangles) correspond to the electronic energy
obtained with the ED (\textit{Dyn}) method at inverse temperatures $\beta $%
=10 and 100 respectively, for $\tau _{sim}$=400.}
\label{fig:figure6}
\end{figure}

The total electronic energy is obtained by integration of $\rho \left(
\omega \right) ,$ and in Fig. 6 its dependence vs. chemical potential $\mu $
is shown, where the full (red squares) and dashed lines (black triangles)
corresponds to the electronic energy obtained with the ED ($Dyn$) method at
inverse temperatures $\beta $=10 and 100 respectively, for $\tau _{sim}$%
=400. For the depicted $\omega -$range it can be observed at low
temperatures two plateaus centered around the eigenvalues $-$4$t_{h}$ and $-$%
3,41421$t_{h}$, obtained after integrating the delta-shaped density of
states.

\begin{figure}[h]
\includegraphics[width=8cm]{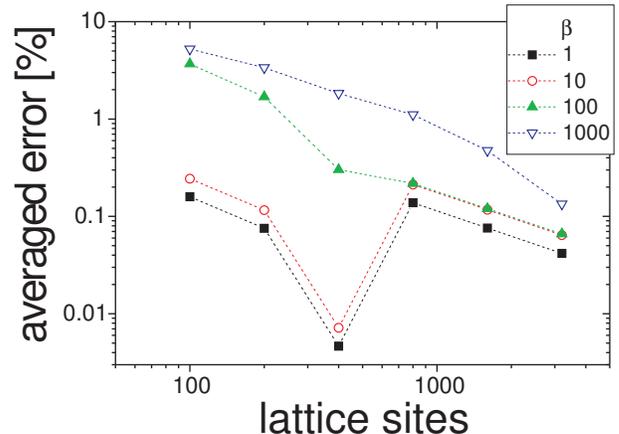}
\caption{Averaged error of electronic energies as a function of $\tau _{sim}$%
, for values of $\beta $=1 (squares), 10 (circles), 100 (up triangles), and
1000 (down triangles).}
\label{fig:figure7}
\end{figure}
We define the relative averaged error of the \textit{Dyn} method in the
electronic energy, for $\mu \in \left[ -2Dt_{h},+2Dt_{h}\right] $ by,

\[
\text{averaged error}=\frac{1}{N_{\tau _{sim}}}\sum_{i=1}^{N_{\tau _{sim}}}%
\frac{\sqrt{\left( E_{ED,i}-E_{Dyn,i}\right) ^{2}}}{E_{ED,i}} 
\]
where the electronic energies $E_{ED,i}$ and $E_{Dyn,i}$ were calculated
using both methods for $N_{\tau _{sim}}$=6400 equispaced frequencies in the
interval $\left[ -2Dt_{h},+2Dt_{h}\right] $, and plotted in Fig. 7 as a
function of $\tau _{sim}$, for values of $\beta $=1 (squares), 10 (circles),
100 (up triangles), and 1000 (down triangles). We can observe that the
averaged error is larger when $\beta $ increases, and decreases when $\tau
_{sim}\,$is increased, but in a non-monotonic way, and it diverges for
values of $\mu $ out of the band. In the present work, the $Dyn$ method will
be used together with the MC technique for filling x=1/2, where the averaged
error is less than 0.1\%. High precision schemes can be achieved for the
case when $\tau _{sim}/\tau _{wf}=L\gg 1$ and $L\lesssim mod(\tau
_{sim},\tau _{wf})$, where $L$ is an integer number and $mod()$ is the
modulus operation.

\section{Case A: one orbital per site and PM configuration}

The case of a random distribution for $\theta _{i}$ and $\varphi _{i}$: $%
\theta _{i}\in \left[ 0,\pi \right] $ and $\varphi _{i}\in \left[ 0,2\pi
\right] $ is relevant in the present model of Eq. (1) as it represents a
particular statistical realization of the paramagnetic (PM) phase, and it
will be used as another test for the $Dyn$ method. In Fig. 8 it is shown the
time dependence of $Im\left[ G(i_{0},i_{0};\tau )\right] $ for the FM (full
black line) and the PM (blue open circles) cases. In the last case, a random 
$i_{0}$ site in a D=2, $N$=6400 sites lattice (80x80) was chosen. We can see
an irregular time dependence, with an average characteristic period longer
than the FM case.

\begin{figure}[h]
\includegraphics[width=8cm]{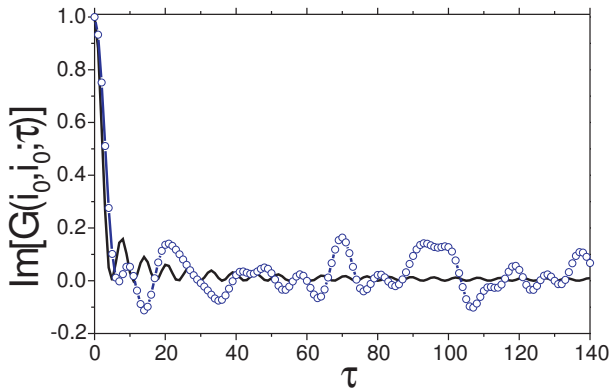}
\caption{Time dependence of $Im\left[ G(i_{0},i_{0};\tau )\right] $ for the
FM (full black line) and the PM (blue open circles) cases, for a
random-picked $i_{0}$ site in the 80x80 square lattice.}
\label{fig:figure8}
\end{figure}

The corresponding PM local electronic density $\rho _{i_{0}}(\omega )$ is
depicted in Fig. 9(a), for the same random-chosen site $i_{0}$. The total
density of states can be obtained after averaging the local density of
states $\rho (\omega )=\frac{1}{N}\sum_{i_{0}=1}^{N}\rho _{i_{0}}(\omega )$,
for the particular random $\left\{ \varphi _{i},\theta _{i}\right\} $
configuration considered (Fig. 9(b), red line). The disorder in the hopping
distribution in the PM phase reduces the FM band-width (Fig. 9(b), black
line) by a factor $1/\sqrt{2}$. The quantity $\rho (\omega )$ can be
obtained analytically for the PM case after considering this reduction of
the FM band-width, but the ED results are not shown here, since the CPU time
needed is about 6 orders magnitude longer that the corresponding using the $%
Dyn$ method.

\begin{figure}[h]
\includegraphics[width=8cm]{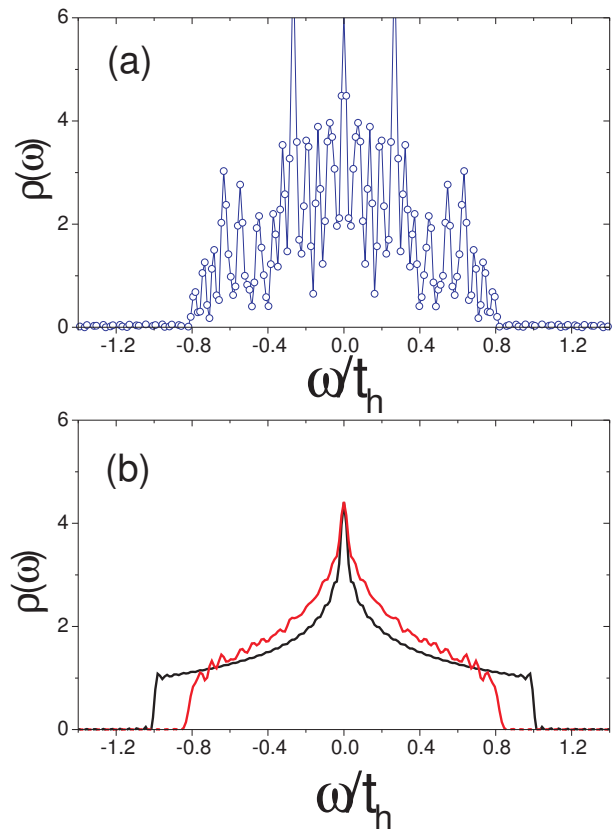}
\caption{(a) Local density of states $\rho _{i_{0}}\left( \omega \right) $
for a random spin configuration $\left\{ \varphi _{i},\theta _{i}\right\} $
on a 80x80 lattice. (b) Total density of states $\rho (\omega )=\frac{1}{N}%
\sum_{i_{0}=1}^{N}\rho _{i_{0}}(\omega )$ ($N$=6400) for the PM (red line)
and the PM (black line) phases. }
\label{fig:figure9}
\end{figure}

As the calculation of $\rho _{i_{0}}(\omega )$ is completely independent of
the site considered, a speed-up of the algorithm should be achievable by
means of a computational $parallel$ implementation.\cite{speedup}

\section{Case B: two orbitals per site and the MC algorithm for ground state}

The full expression of Eq.(1) is taken into account, together with the
consideration of two orbitals per site. The ground state is sampled through
the calculation of the partition function, where the electronic component is
given by:

\begin{equation}
\log Z_{el}=\int_{-\infty }^{+\infty }\rho \left( \omega \right)
(1+e^{-\beta \left( \omega -\mu \right) })d\omega  \label{19}
\end{equation}
obtained for each fixed proposed configuration the angles and displacements
coordinates $\left\{ \theta _{i},\varphi
_{i},u_{x,i},u_{y,i},u_{z,i}\right\} $ for each site, and $\beta $ is the
inverse temperature. The model is analyzed primarily using a classical Monte
Carlo (MC) procedure for the localized spins and phonons, in conjunction
with the ED and $Dyn$ methods for the electronic matrix. This last part of
the process corresponds to the solution of the single-electron problem with
hoppings determined by the localized spin and phonon configuration. The
resulting electronic density is then filled with the number of electrons to
be studied, that is, the simulations are carried out in the canonical
ensemble.

A quantitative comparison of the electronic energies between both methods is
given in Fig. 10, where the total averaged energies per site obtained using
the ED (black triangles) and $Dyn$ (red circles) techniques, as a function
of the parameter $J_{AF}/t$. The Monte Carlo simulations were performed on a
4x4 cluster, at a temperature $T/t$=0.025, $\lambda $=1.0, $\beta _{r}$=2
and filling $x$=1/2 (one particle every two sites). The vertical dashed
lines represents approximately the values of $J_{AF}/t$ where the energy
level crossings occur between the FM, the AF-type CE, and the AF-type G
phases. The phase diagram for D=2 was already obtained using the ED method%
\cite{3}, and the \textit{Dyn} method is in excellent quantitative agreement
with these results.

\begin{figure}[h]
\includegraphics[width=8cm]{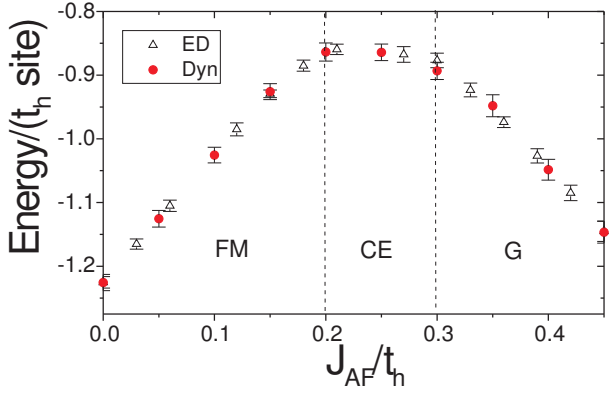}
\caption{Total energies per site obtained for the effective manganites model
(Eq. 1) using the ED (black triangles) and \textit{Dyn} (red circles)
techniques, as a function of the parameter $J_{AF}/t$. The Monte Carlo
simulations were performed on a 4x4 cluster, at a temperature $T/t$=0.025, $%
\lambda $=1.0, $\beta _{r}=2$ and $x$=1/2. The dashed lines represents
approximately the $J_{AF}/t$ where the energy level crossings occur between
the FM, the AF-type CE, and the AF-type G phases.}
\label{fig:figure10}
\end{figure}

\section{Results for D=3}

Novel results for the phase diagram x=1/2, $\lambda $=0.5, D=3 at $T/t_{h}$%
=0.02 were obtained following MC+\textit{Dyn} simulations on a 12x12x12
cluster (3456 orbitals). The maximum number of momenta considered was $%
K_{max}$=100 or 200, and the number of Monte Carlo steps per site (MCS/S)
was typically taken as 5000. Full parallelization of the algorithm was
performed, where tipically 288 cpus where dedicated to compute a single
task. Updates of the spin and phononic $\left\{ \theta _{i},\varphi
_{i},u_{x,i},u_{y,i},u_{z,i}\right\} $ configurations were accepted or
rejected according to the Metropolis algorithm. The simulations start in
most of the cases with random initial configurations, but for the A and CE
phases the convergence is very slow, specially close to the energy
crossovers. A speed up of the convergence was realized by fixing the
corresponding spin configurations, and testing the stability of the proposed
ground state as a function of MCS/S. The magnetic character of the different
ground states were analyzed by means of Spin Structure Factor $S(\mathbf{{q}%
)=}1/N^{\mathbf{2}}\mathbf{\sum_{i,j}\langle {S}_{i}{.S}_{j}\rangle e^{i%
\mathbf{q}{.}({r}_{i}-{r}_{j})}}$. For the 12x12x12 cluster, the $\mathbf{q}$
values considered are: ($l\pi /6$,$k\pi /6$,$m\pi /6$), where $l$, $k$ and $%
m $ are integers values in the interval $0\leq l,k,m\leq 6.$

\begin{figure}[h]
\includegraphics[width=8cm]{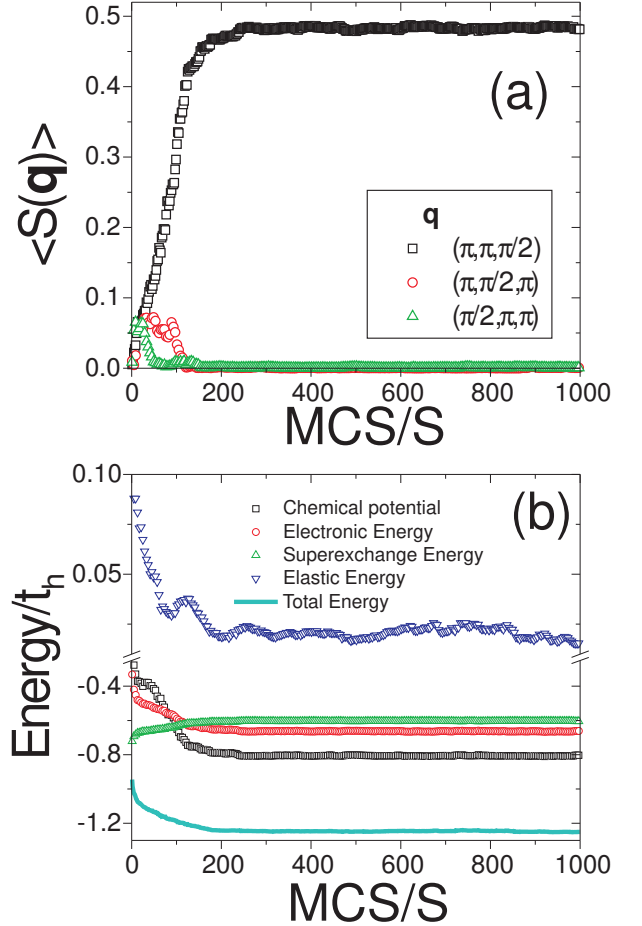}
\caption{(a) $S(\mathbf{{q})}$ vs. MCS/S for $J_{AF}/t_{h}$=0.3, x=1/2, $%
\lambda $=0.5, D=3 at $T/t_{h}$=0.02 on a 12x12x12 cluster and $\mathbf{q}%
=(\pi ,\pi ,\pi /2)$ (black squares), $\mathbf{q}=(\pi ,\pi /2,\pi )$ (red
circles) and $\mathbf{q}=(\pi /2,\pi ,\pi )$ (green triangles). The ground
state corresponding to this setup of parameters is the dimer phase. (b)
Chemical potential $\mu $ (black squares), electronic (red circles),
superexchange (green up triangles), elastic (blue down triangles) and total
energies (cyan line) vs. MCS/S.}
\label{fig:figure11}
\end{figure}
In Fig. 11(a) it is shown $S(\mathbf{{q})}$ vs. MCS/S for $J_{AF}/t_{h}$=0.3
and $\mathbf{q}=(\pi ,\pi ,\pi /2)$ (black squares), $\mathbf{q}=(\pi ,\pi
/2,\pi )$ (red circles) and $\mathbf{q}=(\pi /2,\pi ,\pi )$ (green
triangles). The ground state corresponding to this setup of parameters is
the dimer phase, which consists of pairs spins $\uparrow \uparrow ,$ aligned
antiferromagnetically between them. This phase is evidenced by a single peak 
$S(\mathbf{{q})=}1/2$ at one of these $\mathbf{q}$ values chosen. For MCS/S$%
\lesssim $150 the dimer correlations between neighboring spins are pointing
in all the three spatial directions. For MCS/S$>$200, the system breaks the
spatial isotropy, and all the cluster consist of dimers aligned in the $z$%
-direction, for this particular example, and for MCS/S$>$400 the system has
reached thermal equilibrium. In Fig. 11(b) it is shown the Chemical
potential $\mu $ (black squares), electronic (red circles), superexchange
(green up triangles), elastic (blue down triangles) and total energies (cyan
line) vs. MCS/S. The Chemical potential $\mu $ is calculated
autoconsistently at each Monte Carlo step to fix the electronic density to $%
x=0.5\pm 0.00001$. The electronic energy is given by the first and fourth
terms in Eq. (1), the superexchange energy involves the second term and the
elastic energy is given by the fifth term.

The total energies per site of the ground states are plotted as a function
of the parameter $J_{AF}/t_{h}$ in Fig. 12, under the same conditions used
in Figs. 11. As the $J_{AF}/t_{h}$ parameter is increased, the AF-insulating
character of the ground states obtained increases. Comparing this picture
with the phase diagrams for D=2 (Fig. 10), we notice the appearance of the A
and dimer phases for the case D=3. The A-phase, which consists of FM planes
aligned antiferromagnetically between them, and the FM-phase are degenerate
for D=2, but for the case D=3 their corresponding total energies are
different, i.e. the degeneracy is removed. The dimer phase, which was
obtained in D=1 and 2\cite{4} considering one orbital per site, is not a
ground state in D=2 with two orbitals per site due to the explicitly
breaking of the symmetry between $d_{x^{2}-y^{2}}\ $and $d_{3z^{2}-r^{2}}$
orbitals.

Increasing the insulating parameter $J_{AF}/t_{h}$ the following sequence of
phases is obtained: FM, A, CE, dimer and G%
%
. It is worth to note that, with the exception of the dimer phase, these
phases have been observed experimentally in half-doped manganites,\cite
{tomioka,kuwahara,okimoto,wollan,jirac} following approximately the same
sequence.\cite{kajimoto}

For values of $J_{AF}/t_{h}$ in the interval $\left[ 0,0.125\right] $ the
ground state of the system corresponds to the metallic FM phase. The orbital
configuration of this phase is highly degenerate, but analyzing snapshots at
finite temperature orbital states of the form cos$\left( \Theta \right) \mid
d_{x^{2}-y^{2}}\rangle +$sin$\left( \Theta \right) \mid
d_{3z^{2}-r^{2}}\rangle $ with random values of $\Theta $ in the range $%
0\leq \Theta \leq 2\pi $ are observed, namely the orbital state of this
phase is completely disordered with an homogeneous distribution of the
charges.

\begin{figure}[h]
\includegraphics[width=8cm]{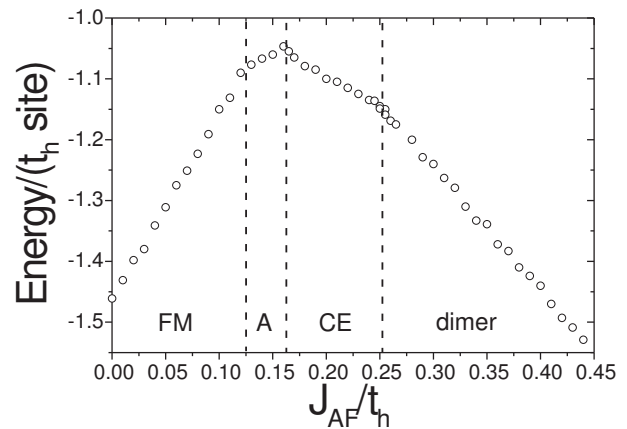}
\caption{Total energies per site of the ground states as a function of the
parameter $J_{AF}/t_{h}$ for the same conditions as in Fig. 11. }
\label{fig:figure12}
\end{figure}

The A-phase is found for values of $J_{AF}/t_{h}$ in the interval $\left[
0.125,0.16\right] $. The orbital distribution corresponds to a majority
occupation of the planar $d_{x^{2}-y^{2}}$ orbitals, which favors a metallic
in-plane conductivity. At the same time this orbital arrangement favors the
AF alignment between neighboring FM\ planes, where the hopping process
inter-planes is strongly suppressed. This phase has been observed in the Pr$%
_{0.5}$Sr$_{0.5}$MnO$_{3}$\cite{tomioka} and Nd$_{0.5}$Sr$_{0.5}$MnO$_{3}$%
\cite{kuwahara} compounds, in this last case in coexistence with the
CE-phase.

For values of $J_{AF}/t_{h}$ in the approximate interval $\left[
0.16,0.26\right] $ the ground state of the system corresponds to the
CE-phase.\cite{okimoto} This magnetic phase was measured experimentally in
the systems La$_{0.5}$Ca$_{0.5}$MnO$_{3}$\cite{wollan}, Nd$_{0.5}$Sr$_{0.5}$%
MnO$_{3}$\cite{kuwahara}, and Pr$_{0.5}$Ca$_{0.5}$MnO$_{3}$\cite{jirac}.
Even for $\lambda $=0, the character of this phase is insulating and was
analyzed in several theoretical works\cite{hottaCE,vbrink,3}, were the
details of the spin, charge and orbital configuration can be seen.\cite
{commentI}

The dimer phase, which is the ground state of the model in $J_{AF}/t_{h}\in
\left[ 026,0.66\right] $ for D=3, was also obtained in D=1 and 2\cite{4}.
The existence of FM dimers (Zener Polarons) along the zigzag chains of the
CE-phase has being observed in Pr$_{0.6}$Ca$_{0.4}$MnO$_{3}$\cite{aladine}
compounds, together with a dimerization of the Mn-Mn distances. More complex
interactions should be included in the present model, like the displacement
of the Mn cations together with the distance dependence of the couplings $%
J_{AF}$ and $t_{h}$ in oder to reproduce this phase$.$ For the pairs $%
\uparrow \uparrow $ directed in the $\xi =\mathbf{x},$ $\mathbf{y}$ or $%
\mathbf{z}$ spatial directions, the orbital configuration of the dimers
phase corresponds to orbitals $d_{3\xi ^{2}-r^{2}}.$

Finally, for values $J_{AF}/t_{h}\gtrsim 0.66$ the insulating $G$ phase is
obtained. The range of parameters where the G and dimer phases are present
corresponds to manganites with very small band-width and sofar has not been
experimentally observed up to date.

\section{Conclusions}

In this work it was described an efficient high-performance algorithm to
calculate the local time-dependent Green's operator for general hermitian,
time-independent, linear differential operator $H$, expressed in matrix
form. The time-dependent local Green's operator is expressed as a finite
series of the Chebyshev polynomials of the normalized $H$ operator,
multiplied by the corresponding Bessel function of the first kind\cite{5}.
This approach, together with the fact that in the local basis no
multiplication of matrices by vectors is needed to keep track of the
time-evolution of an initial state, leads to an algorithm where CPU time and
memory scale linearly with the number of states in the basis of the Hilbert
space.

The total energies of an effective Hamiltonian for manganites were compared
using the Exact Diagonalization technique and the novel \textit{Dyn} method.
The results obtained with both ED and \textit{Dyn} methods were in general
good agreement, and for lattices larger than 289 sites the \textit{Dyn}
method outperforms the ED and TPEM techniques. A parallelization of the
algorithm is possible, with a speed-up factor close to the number of
processors involved. The \textit{Dyn} method could become the method of
choice to expand the present computational limitations set by the ED method,
and in particular to study the model for D=3.

It was shown that it is possible to recover results for the infinite-size
limit\cite{feiguin}, provided the maximum time of the simulation on a
finite-lattice is less than the time that the 'wave front' reaches back to
the site $i_{0}$, $\tau <\tau _{sim}$. The discreteness nature of the
spectra is obtained in the opposite limit, $\tau \gg \tau _{sim}$.

The \textit{Dyn} method used together with the Monte Carlo technique was
used to obtain new results for the manganites phase diagram for D=3 and
half-filling in a 12x12x12 cluster (3456 orbitals). As a function of an
insulating parameter we found the following sequence of ground states: FM,
AF-type A, AF-type CE, dimer and AF-type G. These phases are in remarkable
agreement with experimental results in half-doped manganites,\cite
{tomioka,kuwahara,okimoto,wollan,jirac} which follow approximately the same
sequence.\cite{kajimoto}

\section{Acknowledgements}

We are grateful to A. A. Aligia, M. Kollar, D. Vollhardt, and V. V.
Dobrovitski for helpful discussions and to G. Alvarez for providing the TPEM
code. The CSIT at FSU and the Forschungszentrum at Juelich are also
acknowledged. This work was supported by the Deutsche Forschungsgemeinschaft
through SFB 484 (BN).


\begin{thebibliography}{99}
\bibitem{pick}  That means that we can pick arbitrarily large values for $%
\mid \gamma _{\max }\mid $, but at the same time we will need to calculate
longer time evolutions.
\bibitem{accuracy}  The numerical accuracy of the computer-generated Bessel
function of the first kind is reduced substantially for arguments greater
than $\tau \approx $75. In order to avoid this problem, the initial state
have to be refreshed periodically, for times $\tau \leq $75.
\bibitem{noneed}  It is worthy to remark, that despite the matricial nature
of the operators in the series expansion given by Eq.(9), no matrices or
multiplication of matrices and vectors are needed in the code of the
algorithm. For a more specific example, let's assume we have free spin-less
fermions in a chain of three sites, and $X=t\left[ \mid 1\rangle \langle
2\mid +\mid 2\rangle \langle 3\mid +\mid 3\rangle \langle 1\mid +h.c.\right] 
$ (the parameter $t$ is already normalized) and we want to trace the
evolution of the initial state $\mid i_{0}\rangle =\mid 2\rangle $.
Following Eq.(9), we obtain: $\mid \nu _{0}\rangle =\mid 2\rangle $, $\mid
\nu _{1}\rangle =X\mid \nu _{0}\rangle =t$ $\left[ \mid 1\rangle +\mid
3\rangle \right] $, $\mid \nu _{2}\rangle =X\mid \nu _{1}\rangle -\mid \nu
_{0}\rangle =t^{2}$ $\left[ \mid 2\rangle +\mid 3\rangle +\mid 1\rangle
+\mid 2\rangle \right] -\mid 2\rangle =t^{2}$ $\left[ \mid 3\rangle +\mid
1\rangle \right] +\mid 2\rangle (t^{2}-1)$, and so on.
\bibitem{speedup}  The speed-up factor in this case is close to the number
of procesors, $N_{p}$. Each of the processor must evaluate the described
algorithm starting with a different initial state $\mid i_{0}\rangle $, and
then they must communicate only the fourier-transformed value of $%
G(i_{0},i_{0};\omega )$ to a single processor, which it in turn performs the
summation.
\bibitem{commentI}  The $I$ and $120$ phases\cite{vbrink} have not being
observed as ground states of the present model in 3D, with the exemption of
4x4x4 clusters and for fillings $x\leq $0.45. However it is remarkable that
the energies per site of the phases $CE$ and $I$ are normally as close as
0.02$t_{h}$in clusters 4x4x4, 8x8x8 and 12x12x12.
\end{thebibliography}
\end{document}